\documentstyle[pre,aps]{revtex}

\begin{document}
\draft

%\twocolumn[\hsize\textwidth\columnwidth\hsize\csname @twocolumnfalse\endcsname

\title
{
Structure Factor in the Presence of Shear - an RPA Calculation
}
\author
{
Guy Vinograd and Moshe Schwartz
}
\address
{
School of Physics and Astronomy\\
Tel Aviv University, Ramat Aviv, 69978, Israel
}
\date{-- Feb 2002}
\maketitle
\begin{abstract}
We consider the structure factor of a system of colloidal
particles immersed in a host liquid. Each particle is assumed to be
affected by forces due to other particles, a drag force proportional
to the velocity of the particle relative to the local velocity of 
the fluid and a fluctuating random noise. The effect of the particles
on the velocity field of the host liquid is neglected. The problem is
treated within the RPA approximation which is a widely used tool in 
many body theory. The validity of the RPA result is discussed.
\end{abstract}
\pacs{PACS Numbers: 83.50.A, 82.70.Dd, 64.60.My, 61.20.Ne}
%\vspace{0.5in}
%\hspace{0.5in}
%
%\vskip2pc]
%\narrowtext
%
%\subsection{Introduction}
We consider a system of identical particles 
immersed in a host liquid subject
to an external linear shear flow. Such systems have beed studied
experimentally and numerically for the last two decades 
\cite{ackerson80} - \cite{farr97}. The results of those studies
showed not only distortion of the structure factor due to the 
external shear, but also more dramatic effects like shear induced
ordering and jamming. The theoretical explanations were first 
given in a paper by S.~Hess \cite{hess80}, followed by a number of
other papers that tackled the problem from different angles
\cite{hess86} - \cite{schwartz99}. It seems, however, that in spite
of substantial advances made over the years, a simple and full
theoretical description is still lacking.
In the present paper we will obtain the 
shear dependent structure factor using an RPA approximation employed
on an extension of the exact collective coordinates Fokker-Planck equation
derived recently by Edwards and Schwartz \cite{schwartz02}.
The whole method is based on borrowing
approximations that proved simple and effective in the study of
quantum mechanical many body systems. Such methods have been applied
in the past \cite{schwartz70}, \cite{schwartz79} and much more 
simply recently \cite{guy1} to obtain the Percus-Yevick (PY) \cite{py}
equation for a classical hard sphere system. We hope to produce
the analog of the hard sphere PY equation for the case of shear
flow in the very near future.

Our starting point is an assumption concerning the forces acting on
each particle. We assume that each particle is affected due to forces
applied by the other particles, by a drag force proportional to its
velocity relative to the local velocity of the liquid and by a
fluctuating noise which is assumed to be uncorrelated for
different particles. It is further assumed that the velocity field 
of the liquid is given externally and is unaffected by the motion of
the particles. This assumption will be checked in a future
publictaion. We can say at present, that the main effect we expect
from the inclusion of the modification of the velocity field by the
system of particles is a modification of the interaction between
particles and a modification of the noise. In any case, our present
model is described by the set of Langevin equations 
\begin{equation} \label{eq:langevin}
\gamma [\dot{\vec r_i} - \vec V(\vec r_i)] 
= 
-\nabla_i \sum_{j=1}^N w(\vec r_i- \vec r_j) + \vec \eta_i(t)
,
\end{equation}
where $\vec r_i$ is the radius vector of the $i$'th particle,
 $ \vec V( \vec r) $
is the divergenceless velocity field of the liquid,  $ \gamma $ is the
friction constant, $N$ is the 
number of particles, $w$ is the two
body potential between particles and $ \vec \eta_i(t) $ is a random 
force acting on the $i$'th particle obeying
\begin{equation}
\langle \eta_i^k(t) \rangle = 0
\end{equation}
and
\begin{equation}
\langle \eta_i^k(t) \eta_j^l(t') \rangle 
= 
2\sigma \delta_{ij}\delta_{kl}\delta(t-t')
,
\end{equation}
where $k$ and $l$ denote cartesian components of a vector.
The Langevin equation above leads in a standard way to 
a Fokker-Planck equation for the probability 
distribution of particle coordinates, $ P $
\begin{equation} \label{eq:cartesianFokkerPlanck}
\frac{\partial P}{\partial t} = \frac{1}{\gamma} \sum _{i,k} 
\frac{\partial}{\partial r_{ik}}
\left[
kT \frac{\partial P}{\partial r_{ik}} + 
\frac{\partial U}{\partial r_{ik}}P - 
\gamma V_k(\vec r_i)P
\right]
\equiv LP,
\end{equation}
where \[ U \equiv \frac{1}{2} \sum_{i \neq j} w(|\vec r_i -
\vec  r_j|) \] is the total potential energy, 
and $ kT \equiv \frac{\sigma}{\gamma}$.
Since the particles are identical,
the only relevant physical observables are functionals of 
the density
\begin{equation} \label{eq:densityOperator}
\rho(\vec r) \equiv \sum_{i=1}^N \delta(\vec r - \vec r_i).
\end{equation}
To be specific,
we assume that our $N$ particles are enclosed in a cubic box of 
volume $V$ with periodic boundary conditions. The natural collective
coordinates are the Fourier transforms of the density
\begin{equation} \label{eq:densityOperatorFourierTransform}
\rho_{\vec q} (\{\vec {r_i}\})
\equiv
N^{-\frac{1}{2}}
\int 
\rho(\vec r)
e^{-i \vec q . \vec r}
d^3r
\quad \text{ for } \vec q \neq 0 \text{ with }q_i 
\equiv 
\frac{2 \pi n_i}{V^{\frac{1}{3}}}
.
\end{equation}
The Fokker-Planck equation for the probability distribution of
particle coordinates can be transformed into a Fokker-Planck equation for
the probability distribution of collective coordinates by 
following the procedure outlined by Edwards and Schwartz 
\cite{schwartz02}. We note first, that the probability 
distribution of the
collective coordinates, 
$\widehat{P}(\left\{ \rho_{\vec q} \right\})$, is given by
\begin{equation} \label{eq:cartesianToCollectiveTransformation}
\widehat{P}(\left\{ \rho_{\vec q} \right\})
\equiv
\int 
P(\left\{\vec {r_i} \right\})
\prod_{\vec q \neq 0} 
\delta \left(
\rho_{\vec q} - 
N^{-\frac{1}{2}}
\sum_{j=1}^N  e^{- i \vec q . \vec r_j}
\right)%delta
\prod_{l=1}^Nd^3 r_l
.
\end{equation}
Therefore if 
$ O (\{\vec {r_i}\}) $ 
is an observable that can be
written as $O ( \{ \rho_{\vec q} (\{\vec {r_i}\}) \} ) $, then its
thermodynamic average can be calculated by
\begin{equation} \label{eq:collectiveCoordinatesAverage}
\langle O \rangle
=
\int
O (\{\rho_{\vec q} \})
\widehat{P} (\left\{ \rho_{\vec q} \right\})
\prod_{\vec q \neq 0}
d \rho_{\vec q}
.
\end{equation}
(A similar transformation was used previously for the case of
equilibrium statistical physics \cite{guy1}).
We now obtain the Fokker-Planck equation for
$\widehat{P}(\left\{ \rho_{\vec q} \right\})$
by multiplying
eq.~(\ref{eq:cartesianFokkerPlanck}) by 
\begin{displaymath}
\prod_{\vec q \neq 0} 
\delta \left(
\rho_{\vec q} - 
N^{-\frac{1}{2}}
\sum_{j=1}^N  e^{- i \vec q . \vec r_j}
\right)%delta
,
\end{displaymath}
and integrating over the particle coordinates.
In ref. \cite{schwartz02} that considered a system without
shear, an explicit equation is obtained for
$\widehat{P}(\left\{ \rho_{\vec q} \right\})$.
Therefore, here we need only to introduce the term corresponding
to the shear. Since the Fokker-Planck equation is linear, 
the form we obtain is 
\begin{equation} \label{eq:shearFokkerPlanck}
\frac{\partial \widehat{P}}{\partial t} =
L_0 \widehat{P}
+ i N^{-\frac{1}{2}}
\sum_{\vec q , i, k} q_k 
\frac {\partial}{\partial \rho _{\vec q}}
\left[
\int
e^{-i \vec q . \vec r_i}
V_k (\vec{r_i}) P(\{\vec r_m\})
\prod_{\vec p \neq 0} 
\delta \left(
\rho_{\vec p} - 
N^{-\frac{1}{2}}
\sum_{j=1}^N  e^{- i \vec p . \vec r_j}
\right)%delta
 \prod_{l=1}^N d^3 r_l
\right]
,
\end{equation}
where $L_0$ is the linear operator obtained by Edwards and
Schwartz \cite{schwartz02}.
We do not dwell here on bringing the term due to the shear
for a general divergenceless velocity field to the form
$\Delta L  \widehat{P}$. We consider instead the special
case of linear shear
\begin{equation} \label{eq:linearShear}
\vec V(\vec r) = Cx \hat z
.
\end{equation}
In that case, the last term on the R.H.S of 
eq.~(\ref{eq:shearFokkerPlanck}) becomes
\begin{equation} \label{eq:shearElement}
-C
\sum _{\vec q}
q_3
\frac{\partial \rho _{\vec q}}{\partial q_1}
\frac {\partial \widehat P}{\partial \rho _{\vec q}}
.
\end{equation}
The full form of eq.~(\ref{eq:shearFokkerPlanck}), including
the specific $L_0$ is thus given by
\begin{equation}\label{eq:collectiveCoordinatesFokkerPlanck}
\frac{\partial \widehat P}{\partial t}
=
\frac{kT}{\gamma}
\sum_{\vec q}
\left\{
q^2
\left[
\widehat P +
\rho_{\vec q} \frac {\partial \widehat P}{\partial \rho_{\vec q}} +
\frac{\bar \rho}{kT} w(\vec q) \widehat P
\right]
- \nu
q_3
\frac{\partial \rho _{\vec q}}{\partial q_1}
\frac {\partial \widehat P}{\partial \rho _{\vec q}} +
N^{- \frac{1}{2}}
\sum_{\vec p} \vec q . \vec p
\left[
-\rho_{\vec q + \vec p} 
\frac{\partial ^2 \widehat P}{\partial \rho_{\vec q} \partial \rho_{\vec p}}
+\frac{\bar \rho}{kT} w(\vec p) \rho_{\vec q - \vec p} \rho_{\vec p}
\frac{\partial \widehat P}{\partial \rho_{\vec q}}
\right]
\right\}
,
\end{equation}
where $ \nu \equiv \frac{\gamma C}{kT} $ and
$w(\vec q)$ is the Fourier transform of the inter-particle
potential defined by
\begin{equation} \label{eq:PotentialTransform}
w(\vec q) \equiv \int w(\vec r) e^{-i \vec q  . \vec r} d^3r
.
\end{equation}
We will look for a stationary state of the system, hence the L.H.S vanishes.
In order to keep the discussion simple, we will employ the RPA
approximation on the R.H.S of 
eq.~(\ref{eq:collectiveCoordinatesFokkerPlanck}) by 
discarding all the third order terms except for those where
one of the $\rho$'s is $\rho_0=\sqrt N$. 
(Note that derivatives with respect to $\rho_0$ do not
appear). Thus, we're left with the following equation for the
stationary probability distribution:
\begin{equation} \label{eq:RpaFokkerPlanck}
\sum_{\vec q}
\left\{
q^2
\left[
A^{-1}(\vec q) \widehat P +
A^{-1}(\vec q) \rho_{\vec q} 
\frac {\partial \widehat P}{\partial \rho_{\vec q}} +
\frac{\partial ^2 \widehat P}{\partial \rho_{\vec q} \partial \rho_{- \vec q}}
\right]
- \nu
q_3
\frac{\partial \rho _{\vec q}}{\partial q_1}
\frac {\partial \widehat P}{\partial \rho _{\vec q}}
\right\}
=
0
,
\end{equation}
where $A^{-1}(\vec q) \equiv 1 + \frac{\bar \rho}{kT} w(\vec q) $. 
We try a solution of eq.~(\ref{eq:RpaFokkerPlanck}) of the form
\begin{equation} \label{eq:simplestProbabilityDistribution}
\widehat{P}(\left\{ \rho_{\vec q} \right\})
\equiv
e^{
-\frac{1}{2} \sum_{\vec q \neq 0} S^{-1}(\vec q) 
\rho _{\vec q} \rho _{-\vec q}
}
.
\end{equation}
(We denote the pre factor of $\rho _{\vec q} \rho _{-\vec q}$
in the above Gaussian form by $S^{-1}(\vec q)$ since if such
a form does realy solve eq.~(\ref{eq:RpaFokkerPlanck})
then that pre factor will be, as follows from  
eq.~(\ref{eq:collectiveCoordinatesAverage}), the inverse of
the structure factor we are looking for.)
The combination of equations
(\ref{eq:simplestProbabilityDistribution}) and 
(\ref{eq:RpaFokkerPlanck}) yields the following equation for the
structure factor:
\begin{equation} \label{eq:structureFactor1}
\sum_{\vec q}
\left\{
q^2
\left[
A^{-1}(\vec q) - S^{-1}(\vec q)
-A^{-1}(\vec q) S^{-1}(\vec q) \rho_{\vec q} \rho_{- \vec q}
+ S^{-2}(\vec q) \rho_{\vec q} \rho_{- \vec q}
\right]
+ \nu
q_3
S^{-1}(\vec q)
\frac{\partial \rho_{\vec q}}{\partial q_1}
 \rho_{- \vec q}
\right\}
=
0
.
\end{equation}
The last term in eq.~(\ref{eq:structureFactor1}) can be rewritten as
\begin{equation} \label{eq:beforeIntegrationByParts}
\sum_{\vec q}
\nu
q_3
S^{-1}(\vec q)
\frac{\partial \rho_{\vec q}}{\partial q_1}
\rho_{- \vec q}
=
\sum_{\vec q}
\frac{\nu}{2}
q_3
S^{-1}(\vec q)
\frac{\partial}{\partial q_1}
[ \rho_{\vec q} \rho_{-\vec q} ]
.
\end{equation}
Integrating eq.~(\ref{eq:beforeIntegrationByParts})
by parts and combining the result with the rest of the terms in
eq.~(\ref{eq:structureFactor1}) gives
\begin{equation}
\sum_{\vec q}
\left\{
q^2
\left[
A^{-1}(\vec q) - S^{-1}(\vec q)
\right]
+ 
\left[
- q^2 A^{-1}(\vec q) S^{-1}(\vec q)
+ q^2 S^{-2}(\vec q)
+ \frac{\nu}{2}
q_3
S^{-2}(\vec q)
\frac{\partial  S(\vec q)}{\partial q_1}
\right]
\rho_{\vec q} \rho_{- \vec q}
\right\}
= 0 .
\end{equation}
This equation for the static structure factor should hold for every
collective coordinates configuration, $\left\{ \rho_{\vec q}
\right\}$, hence the following conditions must be obeyed:
\begin{equation}
\sum_{\vec q}
q^2
\left[
A^{-1}(\vec q) - S^{-1}(\vec q)
\right]
=
0
\end{equation}
and
\begin{equation}
1
- \frac {S(\vec q)}{A(\vec q)} 
+ \frac{\nu}{2}
\frac{q_3}{q^2}
\frac{\partial  S(\vec q)}{\partial q_1}
= 0
.
\end{equation}
Although it seems that the fact that there is a redundant condition
here may cause trouble, 
it can be easily verified that any solution of the second condition
also satisfies the first condition.
Viewing the second condition as an ordinary differential equation in $q_1$
leads to a solution which depends on an initial condition. This
condition is
determined by applying the constraint that the structure factor must be
finite for all the $\vec q 's$. The structure factor turns out to be
\begin{equation} \label{eq:Ronis}
S(\vec q)
=
\frac{2}{\nu q_3}
\int_{q_1}^{sign(\nu q_3) \cdot \infty }
dq_1'  q'^2 
e^{ 
-\frac{2}{\nu q_3} 
\int_{q_1}^{q_1'} dq_1''
 \frac{q''^2}{A(q'')}
}
,
\end{equation}
where $\vec q' \equiv (q_1', q_2, q_3)$ and 
 $\vec q'' \equiv (q_1'', q_2, q_3)$.
This is the RPA dependence of the structure factor on the shear rate.
A number of points have to be addressed now.
The structure factor is given here in terms of the Fourier transform
of the inter-particle potential, $w(\vec q)$. 
In the case when $w(\vec q)$ does exist, one may replace
$A^{-1}(\vec q)$ in eq.~(\ref{eq:Ronis}) by
the simplest
RPA approximation for the structure factor in the absence
of shear
\begin{equation} \label{eq:RPAStructureFactor}
S_0^{-1}(\vec q) = 1 + \frac{\bar \rho}{kT} w(\vec q)
.
\end{equation}
Indeed, in the limit of zero shear rate, our calculation 
does yield the well known RPA result (\ref{eq:RPAStructureFactor}).
However, The experience with non
sheared liquids suggests 
that even when $w(\vec q)$ exists, in most
cases such RPA approximations produce bad results. 
A useful alternative is to turn eq.~(\ref{eq:RPAStructureFactor})
around and to calculate an effective potential out of the
externally given structure factor. This procedure actually
replaces $w(\vec r)$ by the Ornstein-Zernike direct correlation
function.
Rewriting eq.~(\ref{eq:Ronis}) using an externally given
$S_0^{-1}(\vec q)$ recovers exactly the
expression given by Ronis \cite{ronis84} 
for the structure factor in the presence of shear.
However, a further study of the Ronis expression reveals
an interesting discrepancy when hard spheres interactions
are considered. The structure factor in the absence of shear,
$S_0(\vec q)$, must have the property that the associated pair
distribution function vanishes within the hard spere diameter.
Further more, $S(\vec q)$, the 
structure factor in the presence of shear, must also have this 
property. Unfortunately, it seems that this property
does not follow from the Ronis expression when inesrting the
external expression for $S_0(\vec q)$ given by
Wertheim \cite{wertheim} - Thiele \cite{thiele} 
into the Ronis expression.
We expect to deal
with this problem in the very near future.

%\newpage    %%%%%%%%%%%%%%%%%%%%%%%%%%%%%%%%%%%%
%\twocolumn  %%%%%%%%%%%%%%%%%%%%%%%%%%%%%%%%%%%%
%\narrowtext %%%%%%%%%%%%%%%%%%%%%%%%%%%%%%%%%%%%
%

\newpage

\end{document}